\documentclass[sn-basic,Numbered, authoryear]{sn-jnl}

\usepackage{graphicx}%
\usepackage{multirow}%
\usepackage{amsmath,amssymb,amsfonts}%
\usepackage{amsthm}%
\usepackage{mathrsfs}%
\usepackage[title]{appendix}%
\usepackage{xcolor}%
\usepackage{textcomp}%
\usepackage{manyfoot}%
\usepackage{booktabs}%
\usepackage{algorithm}%
\usepackage{algorithmicx}%
\usepackage{algpseudocode}%
\usepackage{listings}%
\usepackage{custom_funcs}

\usepackage{orcidlink}
\usepackage[numbers]{natbib}
\usepackage{enumitem}
\usepackage{import}

\begin{document}

\title[Article Title]{ChatGPT for GTFS: Benchmarking LLMs on GTFS Understanding and Retrieval}

\author*[1]{\fnm{Saipraneeth} \sur{Devunuri} \orcidlink{0000-0002-5911-4681}}\email{sd37@illinois.edu} 

\author[1]{\fnm{Shirin} \sur{Qiam} \orcidlink{0000-0002-9720-5656}}\email{sqiam2@illinois.edu}

\author[1]{\fnm{Lewis} \sur{Lehe} \orcidlink{0000-0001-8029-1706}}\email{lehe@illinois.edu}

\affil*[1]{\orgdiv{Department of Civil and Environmental Engineering}, \orgname{University of Illinois at Urbana Champaign}, \orgaddress{\city{Urbana}, \state{Illinois}, \country{USA}}}


\abstract{The General Transit Feed Specification (GTFS) standard for publishing transit data is ubiquitous. GTFS being tabular data, with information spread across different files, necessitates specialized tools or packages to retrieve information. Concurrently, the use of Large Language Models(LLMs) for text and information retrieval is growing. The idea of this research is to see if the current widely adopted LLMs (ChatGPT) are able to understand GTFS and retrieve information from GTFS using natural language instructions \emph{without} explicitly providing information. In this research, we benchmark OpenAI's GPT-3.5-Turbo and GPT-4 LLMs which are the backbone of ChatGPT. ChatGPT demonstrates a reasonable understanding of GTFS by answering 59.7\% (GPT-3.5-Turbo) and 73.3\% (GPT-4) of our multiple-choice questions (MCQ) correctly. Furthermore, we evaluated the LLMs on information extraction tasks using a filtered GTFS feed containing four routes. We found that program synthesis techniques outperformed zero-shot approaches, achieving up to 93\% (90\%) accuracy for simple queries and 61\% (41\%) for complex ones using GPT-4 (GPT-3.5-Turbo).
}

\keywords{GTFS, ChatGPT, Large Language Models, Generative AI, GPT-3.5-Turbo, GPT-4}



\maketitle
        
\section{Introduction}\label{intro}
The General Transit Feed Specification (GTFS) is an Open Data Standard (ODS) designed for publishing transit data and serves as a valuable resource for scientific analysis. GTFS facilitates public access to transit agency data, enabling transparency and fostering collaboration \citep{mchugh2013pioneering}. Since its inception with the Bay Area Rapid Transit Agency (BART) in 2006, GTFS has gained widespread adoption, with over 75\% of transit agencies providing scheduled transit services embracing this standard  \citep{Voulgaris2023Predictors}.

Prominent trip planning applications, including Google Maps, Apple Maps, and Open Trip Planner, rely on GTFS feeds to deliver accurate and reliable information to their users. While GTFS simplifies the process of publishing transit data in a standardized format, retrieving and analyzing GTFS data requires an understanding of established conventions (e.g., ``Required," ``May," ``Must," etc.) and terminological definitions (e.g., ``Leg," ``Journey," ``Field," etc.). The feed specification comprises five mandatory files, along with several optional files, each containing one or more primary keys utilized for data integration. There are also nuances within the GTFS specification that demand careful attention to detail. For instance, the calendar\_dates.txt file can be utilized either independently to publish services on all days or in conjunction with the calendar.txt file to specify exceptions to the regular service schedule. Moreover, there exist certain disparities between the references defined by the standard and the recommended best practices. For instance, while the best practices suggest including the \emph{agency\_id} field in the agency.txt file even when the feed contains only one agency, the standard mandates the inclusion of \emph{agency\_id} only when multiple transit agencies are present. These intricacies highlight the importance of meticulous adherence to the GTFS specifications and understanding the recommended practices to ensure accurate data interpretation and utilization.

For this reason, many researchers and practitioners using GTFS rely on software tools, libraries, or packages that validate and analyze GTFS data. For example, the \emph{gtfs-segments} \citep{gtfs_segments,devunuri2022bus} package specializes in organizing GTFS data in the form of segments and calculating the stop spacings. The R5 Routing engine \citep{noauthor_conveyal_2023} and the R5R package \citep{pereira_r5r_2021} can be used for routing analysis, travel time estimations, and accessibility analysis. 

While such tools greatly simplify certain tasks, they involve their own learning curves and are typically specialized. Therefore, in this study, we explore an alternative that may succeed in making GTFS analysis more accessible: Large Language Models (LLMs). Recent research has explored the ability of LLMs to carry out an array of specialized tasks. Some of these tasks involve answering questions about organized text. \citet{bommarito2022gpt} showed that GPT-3.5 successfully passed the multi-state (MBE) multiple-choice component of the Bar exam, which entails answering intricate legal questions. Subsequently, \citet{katz_gpt-4_2023} reported that GPT-4 surpassed both GPT-3.5 and human performance on the entire Uniform Bar Exam. Other research has investigated the potential of LLMs in medicine \citep{lee_chatgpt_2023,ray_assessing_2023,gilson_how_2023}. In addition to answering questions from reading documents, LLMs have been widely adopted for assistance with writing computer code \citep{sobania_choose_2022,sobania_analysis_2023} and \emph{program synthesis} \citep{chen_evaluating_2021,jain_jigsaw_2022,sobania_analysis_2023b}. Program synthesis is a process in which a computer program is automatically generated to solve a given problem. For example, \citet{khatry_words_2023} demonstrated that LLMs could generate code in Python (Pandas) and SQL from natural language prompts for data manipulation tasks. 

ChatGPT and other LLMs have also seen adoption in the transportation domain. \citet{kim_how_2023} examined how ChatGPT identifies key issues in transportation and proposes potential solutions for them. Their results showed the responses from ChatGPT on issues and solutions were in alignment with the transportation research. The study by \citet{zheng2023chatgpt} showcased the potential usage of LLMs for accident information extraction, report automation, information imputation, and report analysis. Concurrently, the research by \citet{mumtarin2023large} offered a comparative evaluation of different LLMs in extracting salient details from crash narratives. Within the realm of public transportation, the study by \citet{voss2023bus} investigated the LLMs' perspective and understanding of topics such as bus bunching and bus bridging, which are instrumental in discussions related to operational efficiency and connectivity of transit networks.

Despite the growth, applicability, and usage of LLMs in transportation and other fields, the use of LLMs to understand and retrieve information from GTFS data has not been explored. GTFS is a form of tabular data assembled according to publically available standard reference. Therefore, if LLMs can understand the semantics of the data, we hypothesize that they should be able to retrieve information from natural language prompts. To the best of our knowledge, ours is the first study examining and benchmarking the understanding of LLMs on the GTFS file format and feeds. Specifically, this research tries to answer the following research questions: 
\begin{enumerate}
    \item To what extent does generative AI, such as ChatGPT, understand the GTFS?
    \item Can LLMs be used to retrieve information from GTFS files?
    \item Does generating code (i.e. Program Synthesis) aid in information extraction? 
\end{enumerate}

To systematically answer these questions, we designed two distinct benchmarks. The first features 195 multiple-choice questions (MCQs), organized into six categories, and is based on the official GTFS documentation. This aims to probe the LLM's understanding of GTFS concepts and semantics. The second benchmark, containing an additional 88 questions across two categories, tests the LLM's proficiency in GTFS information retrieval. We generated these questions by utilizing a filtered GTFS feed, which includes four routes from the Chicago Transit Authority (CTA). For comparison of LLMS, we benchmarked both the free and `Plus' versions of ChatGPT (the state-of-the-art LLMs) in our evaluations. The free version relies on the GPT-3.5-Turbo, while the `Plus' version is built upon the more advanced and parameter-rich GPT-4. While testing the LLMs, we employ different types of prompts or prompting techniques to extract answers and also understand the reasoning. Specifically, we employ zero-shot (ZS), few-shot (FS), chain-of-thought (CoT), and program synthesis (PS) techniques to design the prompts.

The remainder of the paper is structured as follows:  Section \ref{methods} describes the OpenAI API  and the different prompting techniques such as ZS, FS, CoT, and PS that we employed. Section \ref{understand} presents the ``GTFS Understanding'' benchmark and its results using ZS and CoT approaches. Following this, we present the ``GTFS Retrieval'' benchmark and its results using ZS and PS (with one-shot) prompting techniques in section \ref{retrieval}. Finally, Section \ref{conclusion} concludes and discusses potential future work.

\section{Methodology}\label{methods}
The primary objective of this methodology is to identify the limitations in ChatGPT's comprehension of GTFS data. To achieve this, we aim to identify the category where the LLM performs poorly.  Denote the set of questions as $Q$ and the set of potential prompts as $P$. Note that $P$ here is a \emph{hard} prompt i.e. manually crafted text prompt that requires trial and error. The model's output, which consists of the answers, is represented as $A = f(P, Q)$. Here $f$ denotes the LLM which provides a functional mapping between natural language prompts and responses. To evaluate the performance, we have a ground truth set of correct answers denoted as $S$, and a grading function, denoted as $g$, to assess the accuracy of the model's responses. The grading function scores the responses based on the ground truth $S$. The scores in each category of questions give insight into the understanding of GTFS and the lack thereof.

The secondary objective is to extract information from GTFS. The goal here is to find the most effective prompt, denoted as $P^*$, that maximizes the grading function $g$  for the model's output compared to the ground truth:

\[ P^* = \operatornamewithlimits{argmax}\limits_{P}
    g(f(P,Q),S) \]

In essence, we aim to find the prompt configuration that yields the best alignment between the model's answers and the ground truth, thereby improving the overall understanding of GTFS data by ChatGPT. Specifically, we try zero-shot (ZS), few-shot (one-shot), chain of thought (CoT), and program synthesis (PS) methods by changing the prompts and comparing the results. These techniques along with the OpenAI API and its hyperparameters are discussed in detail below.

\subsection{OpenAI API and Hyperparameters}

OpenAI provides API endpoints to interact with their GPT-3.5 and GPT-4 LLMs, besides the chatbot version - ChatGPT. While using the OpenAI {chat completions API}\footnote{OpenAI Chat Completions API: \url{https://platform.openai.com/docs/api-reference/completions/create} [Accessed 2023-07-29]} endpoints, the prompt is not a single string but a list of messages with certain roles and purposes. We use the \emph{openai} python library\footnote{Open AI Python Library \url{https://github.com/openai/openai-python}} that provides wrappers around the chat completions and other API endpoints. The whole prompt is split into two: ``System" and ``User". The ``System" part ensures that all prompts have a similar context of the question to answer and consists of \emph{instructions} given to the GPT. The ``User" part of the prompt has the actual question that is different from the rest. The prompts used in this paper are inspired by examples provided on OpenAI\footnote{Example Q\&A available at \url{https://platform.openai.com/examples/default-qa} [Accessed 2023-07-29]} website.

For this study, we stick to the \emph{gpt-3.5-turbo} (most capable GPT-3.5 model optimized for chat) and \emph{gpt-4} models, which are susceptible to changes in hyperparameters. Besides the prompts, the hyperparameters are the only available features used for tuning the LLM's response. For this study, we use the following key hyperparameters available in the chat completions API:

\begin{itemize}
    \item \emph{temperature}: The temperature parameter (range [0,2]) controls the level of randomness in the sampling. A value of \emph{zero} produces deterministic output, while higher values introduce more randomness. To ensure the reproducibility of results, we set the temperature to \emph{zero}.
    \item \emph{top\_p}: This is an alternative to the temperature parameter that uses nucleus sampling probability. The documentation recommends tuning either temperature or top\_p. Therefore, we use the default value of `1.'
    \item \emph{max\_tokens}: This parameter limits the maximum number of tokens in the output. For multiple-choice type questions, we set max\_tokens to `2'. For question-answer type questions, we set the token limit to the context length of the \emph{gpt-3.5-turbo} model at 16,000 tokens.
\end{itemize}

Besides these, there are other hyperparameters such as `best\_of', `presence\_penalty', and `frequency\_penalty', which we set to their default values in our evaluations.

\subsection{Zero-Shot and Few-Shot Learning}

Zero-shot (ZS) prompting is a technique that allows Large Language Models (LLMs) to retrieve information from data \emph{without} specific training or examples on the domain. The model is only given the instructions in natural language. \citet{kojima2023large} showed that LLMs are zero-shot reasoners and can solve a variety of reasoning problems without fine-tuning or sample engineering.

Since GTFS is a standardized format for publishing transit schedules, the advantage of ZS prompting in GTFS information retrieval lies in its flexibility and adaptability. It allows LLMs to handle a wide range of queries without the need for explicit domain-specific training. This makes it easier to integrate GTFS information retrieval capabilities into various applications, as LLMs can leverage their pre-trained knowledge and generalize to different transit systems and datasets. Figure \ref{fig:mcq_q1} shows an example ZS prompt and response.

\begin{figure}[!h]
    \centering
    \begin{subfigure}{\textwidth}
    \import{}{figures/mcq_questions_prompt.tex}
    \caption{\emph{Q1} is an example of an original MCQ question in the `GTFS Understanding' benchmark }
    \label{fig:mcq_q1a}
    \end{subfigure}
    \begin{subfigure}{\textwidth}
    \import{}{figures/mcq_questions_prompt_variant.tex}
    \caption{The question \emph{Q1.c} is a variant of \emph{Q1} where the option `c' in the original question is replaced with `None of these.'}
    \label{fig:mcq_q1b}
    \end{subfigure}
    \caption{Example questions and answers using zero-shot (ZS) technique on GPT-3.5-Turbo. The prompt consists of system and user portions represented by red and green blocks.  The blue block corresponds to the assistant's (GPT-3.5-
Turbo here) response.}
    \label{fig:mcq_q1}
\end{figure}

\begin{figure}
    \centering
    \import{}{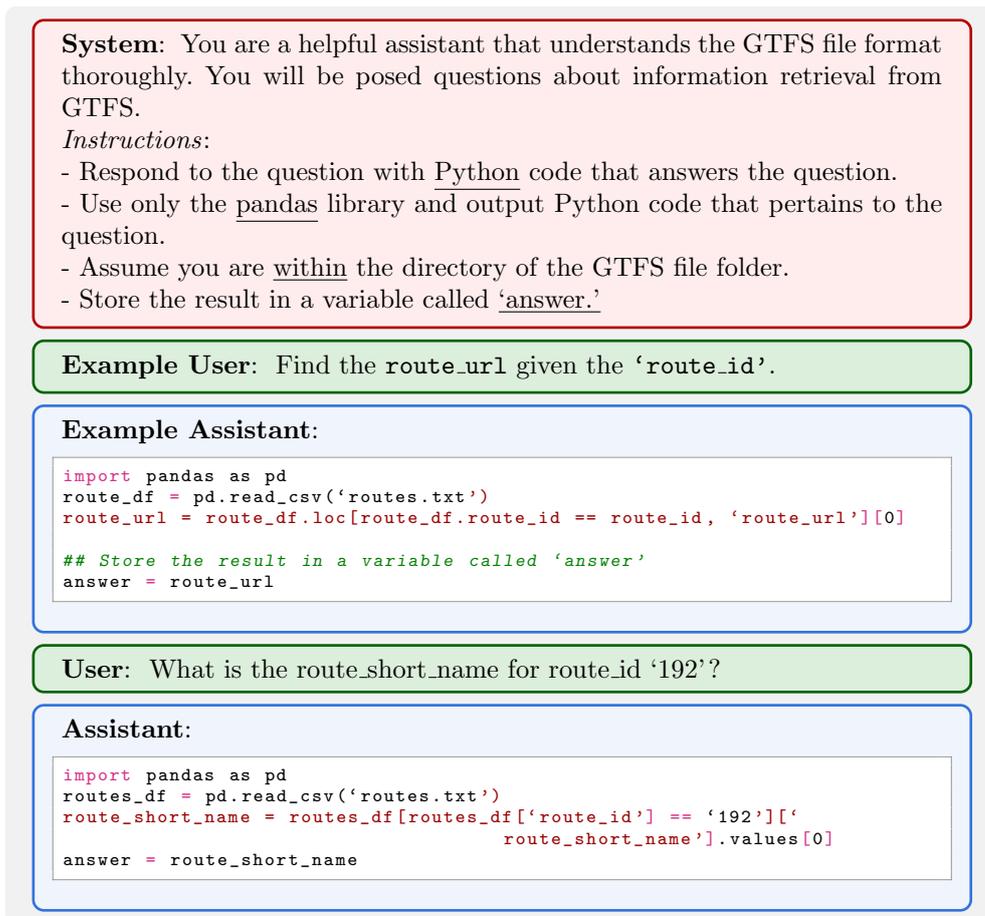}
    \caption{Sample prompt for program synthesis (PS) with the system and user prompts along with assistant output. The `Example User' and `Example Assistant' are used for few-shot (FS) prompting. The response is generated by the GPT-3.5-Turbo model.}
    \label{fig:few-shot}
\end{figure}

Few-shot (FS) prompting is similar to ZS prompting but provides few demonstrations to the model that helps the model respond in a fashion more tuned to the user preference. \citet{brown_language_2020} demonstrated that FS prompting results in better accuracy than ZS by using in-context information. In this paper, we stick to a one-shot setting where we provide a task and a single demonstration of how to solve the task. Figure \ref{fig:few-shot} shows an example of FS questioning.

It is important to note that ZS or FS prompting has its limitations. Since the LLM has not received direct training on GTFS data, its performance may not be as accurate or precise as models specifically trained on GTFS datasets. The level of understanding and retrieval accuracy may vary depending on the complexity of the queries and the available contextual information.

\subsection{Chain of Thought (CoT)}
Chain of Thought (CoT) prompting is a method aimed at eliciting multi-step reasoning from Language Models (LLMs). By guiding the model through a coherent series of steps, CoT encourages it to arrive at an answer systematically. \citet{wei_chain--thought_2023} demonstrated that CoT substantially enhances the reasoning capabilities of LLMs. One of the key benefits of CoT is its ability to provide insights into the thought process employed by the model when answering questions, regardless of whether the LLM's response is accurate or not. A chain of thought could be instantiated by adding something as simple as ``Provide a step-by-step explanation" to the prompt.

\subsection{Program Synthesis (PS)}
Program synthesis (PS) evaluates the understanding of Language Models (LLMs) by generating code that can reproduce the result/answer. This evaluation method provides a more comprehensive assessment of the model's abilities as it requires the LLM to not only interpret the textual input accurately but also produce correct and functional code that fulfills the given requirements. Since most GTFS is tabular data, the pandas \citep{mckinney-proc-scipy-2010} library can be used to generate the code. \citet{khatry_words_2023} applied LLMs to generate code in pandas and SQL.

Additionally, program synthesis, similar to CoT, fosters the interpretability and explainability of LLMs. Since the generated code is executable and traceable, it becomes easier to diagnose the LLM's behavior and identify potential errors or misconceptions. Besides, the code generated can be generalized by asking generalized questions that are not particular to a specific GTFS. This makes it possible to bypass the context length limitation and extract information without the input of GTFS files.





\section{GTFS Understanding}\label{understand}

The GTFS specification has evolved into an intricate Open Data Standard, through changes over the years.  This transformation is driven by its broadening scope and increased adoption across the world by transit agencies. Beyond its original intent of serving as timetable data, the specification now encapsulates information on other aspects surrounding transit such as stop accessibility, pathways, bike facilities, and fare structures. To accommodate the evolving transit service offerings, the specification also added new files and attributes. For example, the file `frequencies.txt` is used to support headway-based schedules, and the attributes `continuous\_pickup' and `continuous\_dropoff' accommodate services where riders can board or alight a vehicle anywhere along the route. As of May 2020\footnote{As of Oct 2023, the changes on May 25, 2020, were the last major change to the GTFS that added three additional files. This change is before the Sep 2021 cutoff for ChatGPT}, the GTFS specification comprises 24 files: 5 mandatory, 3 conditionally required, and 16 optional. Collectively, these files include over 120 unique attributes, either conveying information or facilitating relational mapping between files. While some attributes share the data types and information they represent, certain categorical attributes may further bifurcate what the attribute represents. Moreover, the attributes utilize abstractions for concise data representation, further adding to the complexity.

Given this complexity, understanding GTFS involves a grasp of specification conventions, file structures, inter-file relationships, attributes, and their respective data types \& abstractions. To evaluate the depth of the LLM's understanding of GTFS, we created a benchmark that consists of a diverse array of questionnaires encompassing this complex structure. These questions are based on key parts of the GTFS documentation, transformed into a multiple-choice format. For this study, we stick to the \emph{static} version of GTFS (also termed as GTFS Schedule) alone. However, the same methodology could be applied to the extensions of GTFS such as GTFS-Realtime and GTFS-Flex.

\subsection{GTFS Understanding Benchmark}
 
For our benchmark of GTFS understanding, we employ 195 original multiple-choice questions (MCQs) with single correct answers that were meticulously crafted and span across the six question categories. The benchmarking dataset with all questionnaires is made available to the public (see Section \ref{data_avail} for more info). The breakdown of the number of questions in each category is presented in Table \ref{tab:dataset}. We derived these questions and categories using the official GTFS Schedule documentation available at \url{https://gtfs.org/schedule/reference} as reference. 

 \begin{table}[h]
    \centering
    \caption{GTFS Understanding Benchmarking dataset questionnaire and their categories}
    \label{tab:dataset}
    \begin{tabular}{|cc|}
\hline
\multicolumn{1}{|c|}{\textbf{Question Type}} & \multicolumn{1}{c|}{\textbf{Number of Questions}} \\ \hline
\multicolumn{1}{|l|}{Term Definitions}      & 14   \\ \hline
\multicolumn{1}{|l|}{Common Reasoning}      & 28   \\ \hline
\multicolumn{1}{|l|}{File Structure}        & 17   \\ \hline
\multicolumn{1}{|l|}{Attribute Mapping}     & 32   \\ \hline
\multicolumn{1}{|l|}{Data Structure}        & 30   \\ \hline
\multicolumn{1}{|l|}{Categorical Mapping}   & 74   \\ \hline
\multicolumn{1}{|c|}{\textbf{Total}}   & \textbf{195}   \\ \hline
\end{tabular}
\end{table}

Here is the explanation for each category within this benchmark along with example questions:
\begin{itemize}
    \item \emph{Term Definitions:} GTFS uses terms such as ``Required", ``May", ``Service Day", etc. that are document specific. These are important to understand the file structures and attribute definitions. Examples: 
    \begin{itemize}
    
        \item \texttt{What is a record in GTFS? a) A basic data structure representing a service day b) A collection of field values describing a single entity c) A set of files defining transit information d) A unique identifier for a transit agency}
        \item \texttt{What is a field in GTFS? a) A property of a dataset b) A specific date for transit service c) A basic data structure representing a service day d) A property of an object or entity}
    \end{itemize}

    \item \emph{Common Reasoning:} These questions test the basic knowledge of GTFS such as the abbreviation, the usage, and the purpose of different files present in GTFS. Examples:
    \begin{itemize}
        \item \texttt{What is the purpose of the "agency.txt" file in GTFS? a) It provides information about individual transit stops. b) It contains details about the routes and their associated stops. c) It specifies the frequency of trips. d) It provides information about the transit agency or operator.}
        \item \texttt{Can a GTFS feed include information about multiple modes of transportation? a) Yes, it can. b) No, it can only include information about buses. c) No, it can only include information about trains. d) No, GTFS feeds are specific to a single mode of transportation.}
    \end{itemize}
    
    \item \emph{File Structure:} The GTFS specification has 8 files that are either required or conditionally required and several other files that are optional. These questions test if the GPT engine is able to find the corresponding file based on the general context. Examples:
    \begin{itemize}
        \item \texttt{How is frequency-based service represented in GTFS? a) Through the "frequency.txt" file. b) By using the "calendar.txt" file. c) By specifying trip times in the "stop-times.txt" file. d) Frequency-based service is not supported in GTFS.}
        \item \texttt{What is the purpose of the "levels.txt" file in GTFS? a) It provides information about the levels of fare zones. b) It contains details about the levels or floors of transit stations. c) The "level.txt" file is not applicable in GTFS. d) It specifies the elevation or altitude of transit stops.}
    \end{itemize}
    
    \item \emph{Attribute Mapping:} GTFS has several attributes spread across different files. The questions test if the GPT understands which attributes are present in which files. Examples:
    \begin{itemize}
        \item \texttt{Which file contains the route\_type attribute in GTFS? a) stops.txt b) routes.txt c) trips.txt d) shapes.txt}
        \item \texttt{In which file can you find the service\_id attribute in GTFS? a) stops.txt b) calendar.txt c) shapes.txt d) routes.txt}
    \end{itemize}
    
    \item \emph{Data Structure:} GTFS standard uses several data types that are unique to the specification. For example, the color code for all routes should be a Hex code. These questions test if the attribute data types are correctly identified. Examples:
    \begin{itemize}
        \item \texttt{What data type is used for the date attribute in GTFS? a) Date (YYYYMMDD) b) Enum c) Text representation of the date (e.g., "September 13, 2018") d) Integer (e.g., 20180913)}
        \item \texttt{How is the currency\_amount attribute represented in GTFS? a) Decimal value (ISO 4217) b) Nearest \$5 multiple c) Integer d) Text (e.g., "\$50.00")}
    \end{itemize}
    
    \item \emph{Categorical Mapping:} In order to concisely represent data in attributes, the GTFS feed uses categorical variables to represent data. For example, the `route\_type' `3' corresponds to a `bus' and a `location\_type' `0' corresponds to a `stop'. Examples:
    \begin{itemize}
        \item \texttt{What does "route-type" 6 indicate in the "routes.txt" file? a) Tram, Streetcar, Light rail b) Subway, Metro c) Aerial lift, suspended cable car d) Funicular}
        \item \texttt{Which value in the "location\_type" field of the "stops.txt" file represents a generic marker or point used for reference on a map? a) 0 b) 1 c) 2 d) 3}
    \end{itemize}
\end{itemize}

In evaluating the LLM's performance on MCQs, the model selects the answer (choice) with the highest probability for each question and output that \emph {without} the need for any explanation. An example of this question-answer process is depicted in Figure \ref{fig:mcq_q1}. Although the LLM may always choose the correct answer when it is present, the LLM could opt for an alternate option when the correct choice is missing. To check the LLM's robustness, we generate an augmented question set by creating variations of the original questions. Specifically, each original answer choice denoted as `a', `b', `c', and `d'—is replaced one at a time with the phrase `None of these,' resulting in additional 780 ($195 \times 4$) variant questions and a total of 975 questions in the augmented dataset. Figure \ref{fig:mcq_q1} shows an example of an augmented question where option `c' in the original question is replaced with `None of these.' The augmentation aims to evaluate how well the LLM can adapt to scenarios where the correct answer is removed.

\subsection{Results}
\begin{figure}[!h]
    \centering
    \includegraphics[width=0.95\textwidth]{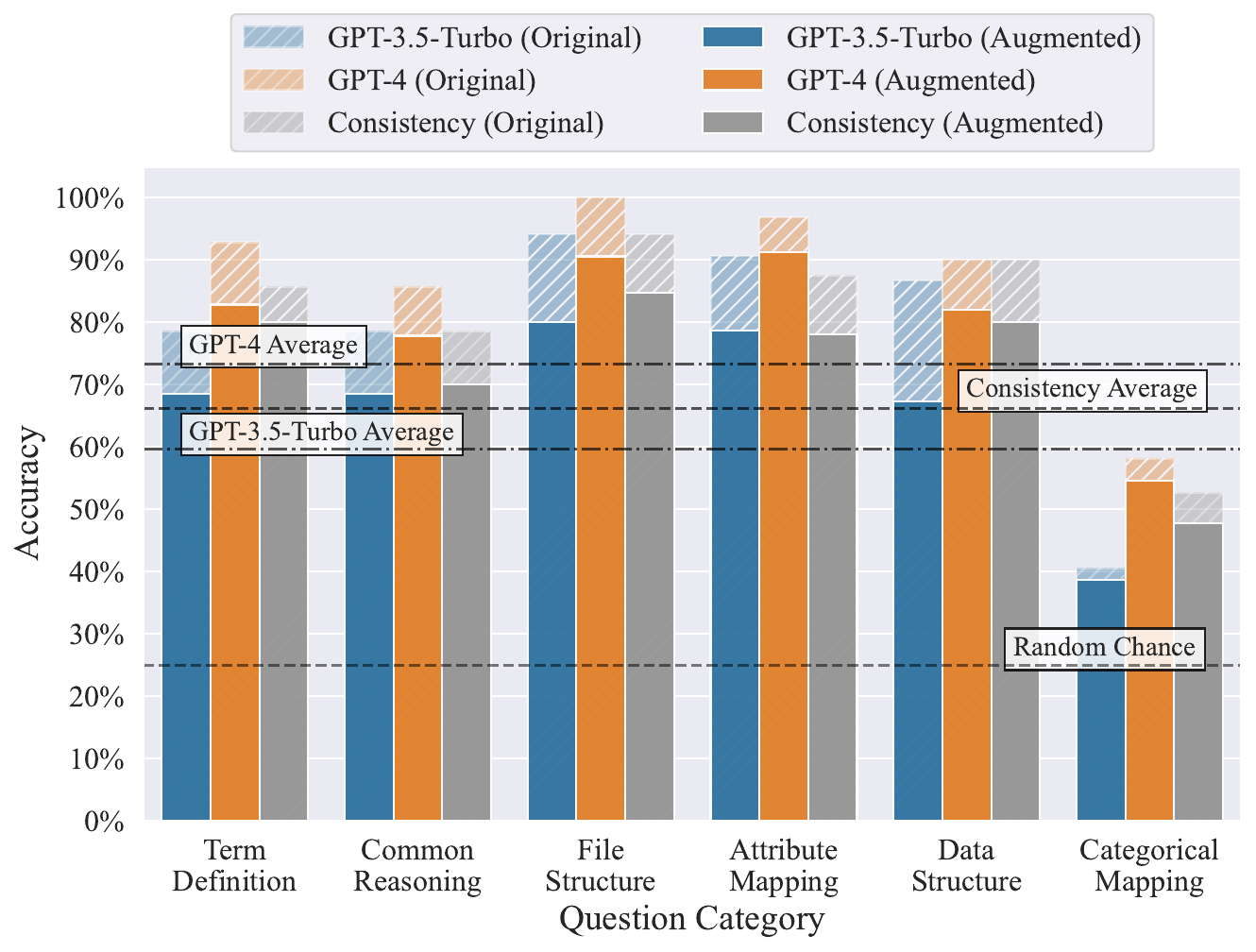}
    \caption{Summary of performance by question category for GPT-3.5-Turbo and GPT-4 on GTFS understanding. The dashed bars represent the original data and the solid bars represent the augmented data.}
    \label{fig:gpt_understand}
\end{figure}

In this study, we benchmarked both GPT-3.5-Turbo and GPT-4 on the original and augmented `GTFS Understanding' benchmark. Using the ZS technique, the LLM attempts to answer the questions without any examples provided to it. The accuracy of ZS on different categories of questions for both GPT-3.5-Turbo and GPT-4, along with their consistency, are shown in Figure \ref{fig:gpt_understand}. Consistency in this context refers to how often both LLMs simultaneously succeed or fail to provide the correct answer. The accuracy and consistency across both LLMs and all categories are higher on the original dataset (shown in dashed) than on the augmented dataset (shown in solid). This indicates that LLMs are not robust to option substitution. The discussions in the remainder of the paper stick to the augmented dataset alone. The overall accuracy across all six categories is 59.7\% for GPT-3.5-Turbo and 73.3\% for GPT-4. In general, these two LLMs display a 66.3\% consistency in their responses to the questions. Overall, GPT-4 performs better than GPT-3.5-Turbo, with above 90\%  accuracy in ``File Structure" and ``Attribute Mapping" type questions; above 80\% in ``Term Definitions" and ``Data Structure"; and below 80\% in ``Common Reasoning" and ``Categorical Mapping" categories. The GPT-3.5-Turbo has around 70\% accuracy for all but the ``Categorical Mapping" category, which has the worst accuracy for both LLMs. Although GPT-4 demonstrates improved accuracy for the ``Categorical Mapping" category, $\approx$15\% higher than GPT-3.5-Turbo, both models encounter challenges in this particular category and are least consistent.

Since LLMs embed words in latent space, the embedding for categorical variables could be similar, making it difficult for the LLMs to distinguish and easier to hallucinate. Take, for example, the categories `Rail' and `Light Rail' in the `route\_type' attribute. The words `Rail' and `Light Rail' have a cosine similarity \footnote{Cosine similarity measures the similarity between two vectors by measuring the cosine of the angle between the vectors. Cosine similarity close to 1 represents similar vectors} of  0.88 with `text-embedding-ada-002' embedding from OpenAI. However, in the context of GTFS, both have different meanings. The `Rail' has `route\_type' of 2, and `Light Rail' has  `route\_type' of 0. Figure \ref{fig:categorical_map} in the Appendix shows an example prompt and response where the GPT-3.5-Turbo fails to answer questions related to categorical mapping. When posed the same question with and without CoT (Figure \ref{fig:categorical_map_ZS} \& \ref{fig:categorical_map_CoT}), we obtained different results. However, both the results were incorrect as the \emph{route\_type} for bus routes is `3'. From the CoT explanation, it is evident that the LLM understands what the `route\_type' purpose is. However, it fails to understand the mapping of categories in it.

\section{Information Retrieval from GTFS}\label{retrieval}

Existing literature has demonstrated that LLMs can extract information from structured tabular data \citep{bai_schema-driven_2023, zhuang_toolqa_2023, khatry_words_2023}. This benchmark evaluates how well LLMs could perform this task on GTFS tabular data. A total of 88 questions were generated of which some are specialized to a particular GTFS feed and others more generally applicable. For questions pertaining to a GTFS feed, we used a filtered feed from the Chicago Transit Authority (CTA) for illustrative purposes. However, these questions can be adapted to any other GTFS feed by replacing the values for attributes like `stop\_id', `trip\_id', `route\_id', etc. in the particular feed, with the loss of generality. 
\subsection{GTFS Retrieval Benchmark}
\begin{figure}[!h]
    \centering
    \begin{subfigure}{0.45\textwidth}
        \centering
        \includegraphics[height=1.4\textwidth]{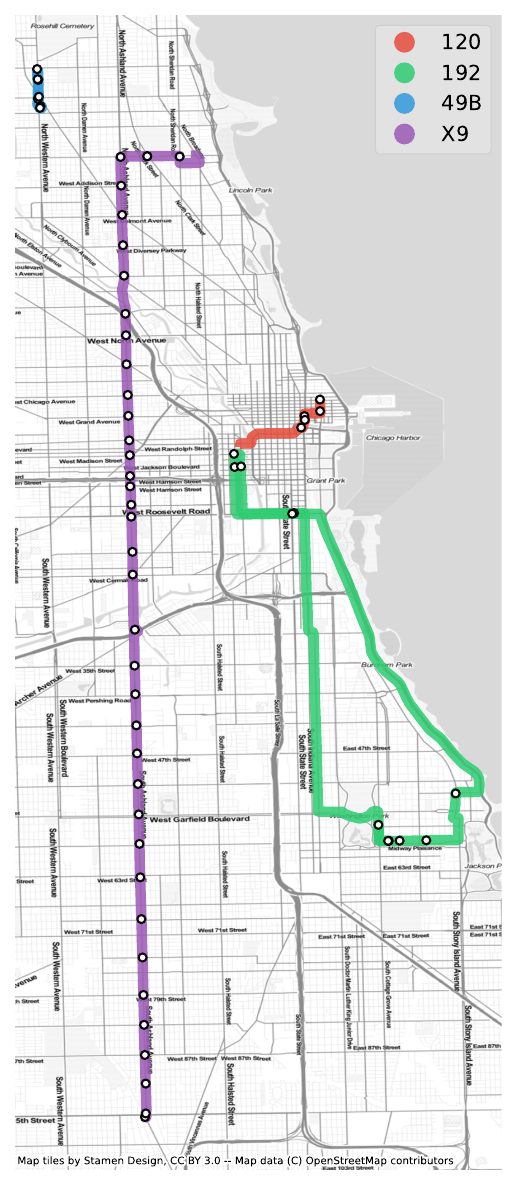}
        \caption{Filtered Network}
        \label{fig:segments_filtered}
    \end{subfigure}
    \caption{Bus and metro network of Chicago Transit Authority (CTA) generated using \emph{gtfs-segments} python library}
\end{figure}

The `GTFS Retrieval' benchmark employs a question-answer (QA) format, where no options are given and the LLM is supposed to give a single, correct answer. To prepare the questionnaire, we used the CTA GTFS feed published on May 3, 2023\footnote{The CTA feed is available for download at \url{https://transitfeeds.com/p/chicago-transit-authority/165/20230503}[Accessed 2023-07-29]}. The full feed included data on 133 routes with 125 bus and 8 metro routes. However, LLMs have limited \emph{context length}: a metric for the number of \emph{tokens} the LLM can process at once. The GPT-3.5-Turbo and GPT-4 have a maximum context length of 16,385 and 32,768 tokens respectively. The full GTFS feed is much larger than either LLM can accept, so we trim the dataset to just four bus routes (`49B', `120', `192', and `X9') and nine trips on these routes. These routes have 63 unique stops. Figure \ref{fig:segments_filtered} shows the map of the filtered dataset.  

These questions in this benchmark range from basic lookup (single or multiple files) to performing data manipulations by the LLM. These include common data manipulation techniques like filtering, sorting, grouping, and joining. We divide the  questions into two categories:

\begin{itemize}
    \item \emph{Simple}: These questions are based on simple lookups within the same file or two different files (using relational keys) within GTFS. Examples:
    \begin{itemize}
        \item \texttt{What is the `stop\_sequence' for stop\_id `5024'?}
        \item \texttt{How many trips visit the stop with `stop\_id' 15164?}
        \item \texttt{How many trips have `direction\_id' as 0?}
        \item \texttt{How many unique `stop\_id' values are there for `route\_id' X9 in the GTFS data?}
        \item \texttt{How many different `direction\_id' are there for `route\_id' 192 in the GTFS data?}
    \end{itemize}

    \item \emph{Complex}: These questions need multiple files to extract information, require a deeper understanding, and could be open-ended. Examples:
    \begin{itemize}
        \item \texttt{What is the `trip\_id' of the trip with the longest duration for the route with `route\_id' 192?}
        \item \texttt{How many trips have a `departure\_time' between 7:00 AM and 9:00 AM on weekdays (Monday to Friday)?}
        \item \texttt{What is the `route\_long\_name' for the route with the most wheelchair-accessible stops?}
        \item \texttt{What is the earliest `departure\_time' for the trips with `service\_id' 65515 on the date 2023-07-05?}
        \item \texttt{What are the `trip\_ids' that have a `shape\_dist\_traveled' greater than 10Km?}
    \end{itemize}
    
\end{itemize}
\subsection{Results}
\begin{figure}[!h]
    \centering
    \import{}{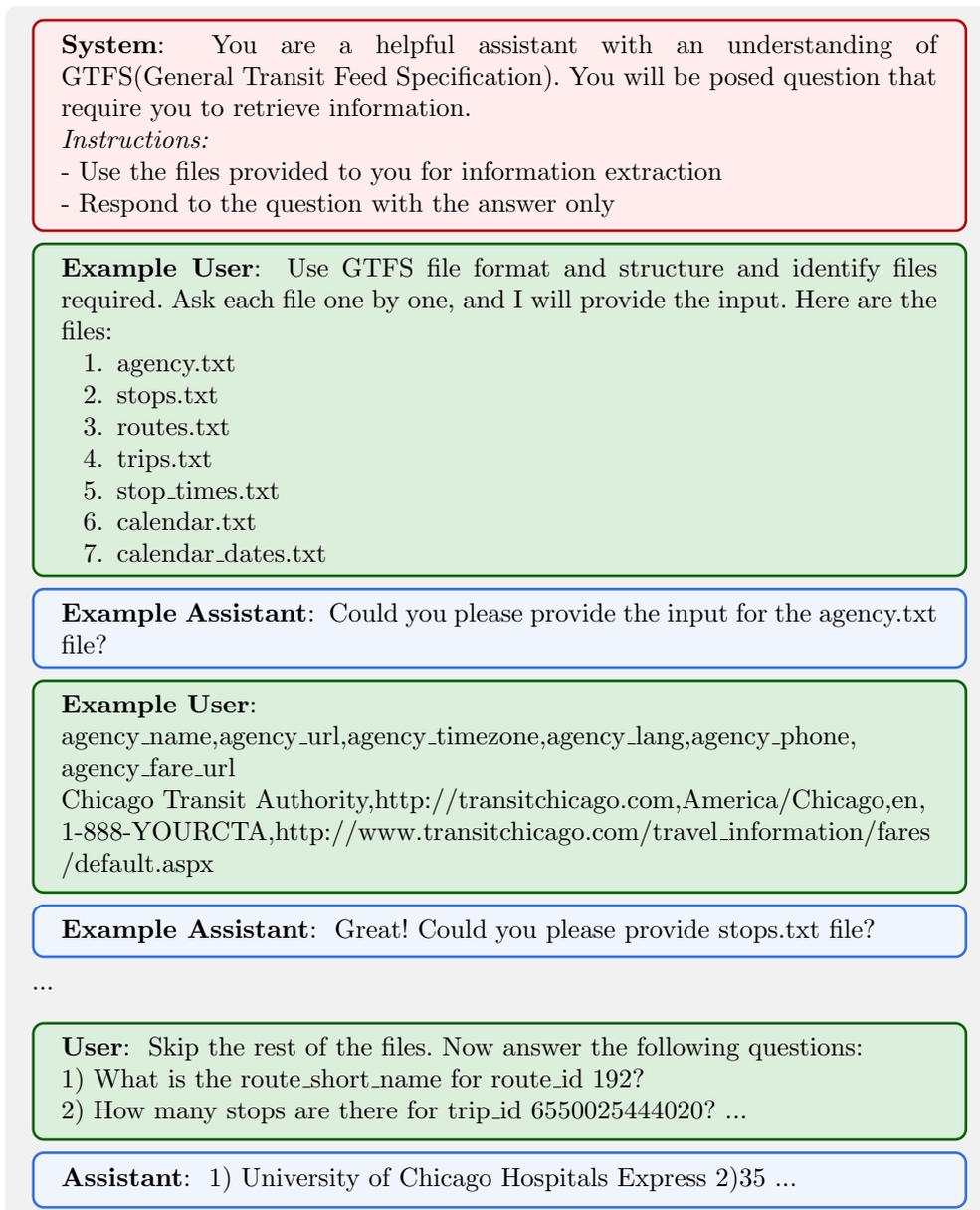}
    \caption{Example Prompt and Response for Zero-shot GTFS Information Retrieval. The `Example User' and `Example Assistant' are used to imitate a conversation. The response is generated by the GPT-3.5-Turbo model.}
    \label{fig:info_input}
\end{figure}

Similar to testing the understanding of GTFS, we pose questions to the LLM to see its capabilities in information retrieval. A total of 88 questions were posed with 47 simple and 41 complex questions. As ChatGPT performance on ``Categorical Mapping" questions is poor, we explicitly mention the value for the categorical variable. We use zero-shot (ZS) and program synthesis (PS) prompting to generate and evaluate the responses. To pose these questions using ZS, we first input GTFS data using a series of prompts. Figure \ref{fig:info_input} gives an example of setting up ZS prompting. Here, we use the `Example User' and `Example Assistant' to imitate a conversation thread where the LLM asks for all the required GTFS files from the user in a step-by-step fashion. Once, all files are input in this fashion, the actual questions are posed. For PS, we use one-shot learning as default. The LLM was able to better generalize the way to output code when provided with an illustration. Figure \ref{fig:few-shot} shows an example prompt for PS with a one-shot example. The code generated by the LLM is later executed in a code interpreter to obtain the results.

\begin{figure}[!h]
    \centering
    \begin{subfigure}{0.60\textwidth}
        \includegraphics[width=\textwidth]{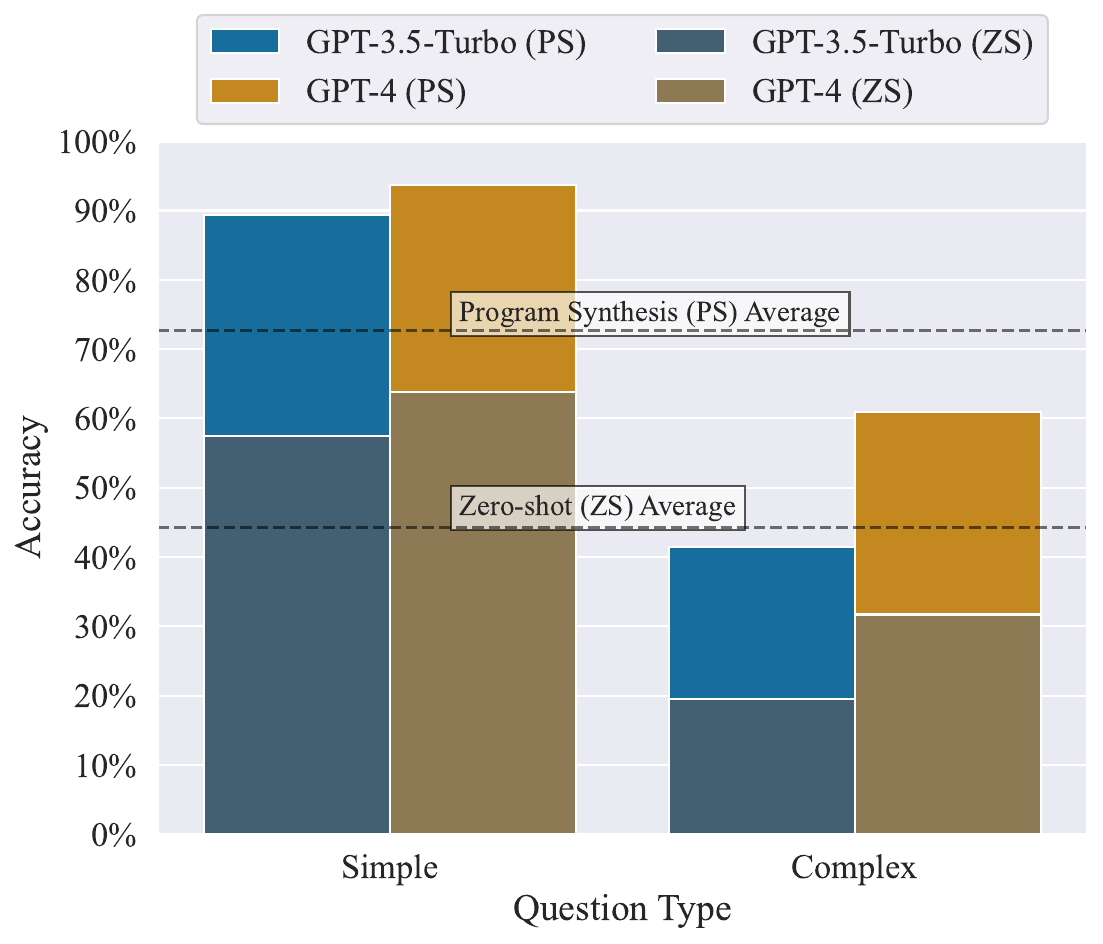}
        \caption{Accuracy of zero-shot and program synthesis}
        \label{fig:accuracy_code}
    \end{subfigure}
    \begin{subfigure}{0.355\textwidth}
        \includegraphics[width=\textwidth]{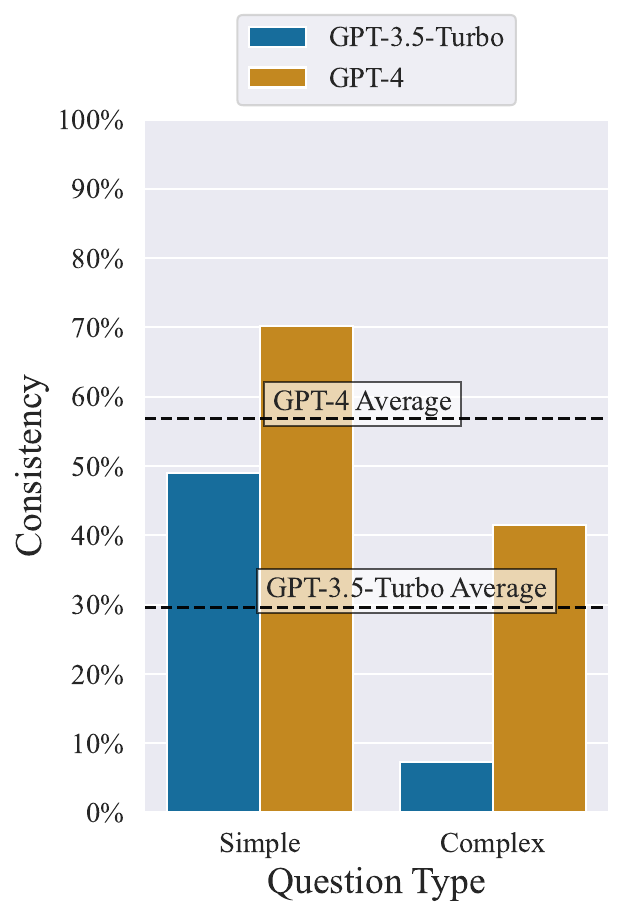}
        \caption{Consistency in outputs}
        \label{fig:consistency_code}
    \end{subfigure}
    \caption{Summary of performance by question type for GPT-3.5-Turbo and GPT-4 on GTFS Retrieval benchmark}
    \label{fig:gpt_retreival}
\end{figure}

The results (see Figure \ref{fig:gpt_retreival}) show that the performance of program synthesis($\approx73$\% accuracy) is significantly better than the zero-shot($\approx44$\% accuracy) for both simple and complex levels of problems. This difference could be arising because of two reasons. With zero-shot, the LLM is unable to make the joins or connections between multiple files and instead hallucinates output or intermediate steps. The second reason would be that similar to CoT, program synthesis makes the LLM think in a step-by-step fashion thereby having fewer logical fallacies. The GPT-4 performs better than GPT-3.5-Turbo across simple and complex categories and in both ZS \& PS techniques. Although GPT-4 improvement is significant in complex questions, both GPT-3.5-Turbo and GPT-4 still struggle with complex questions. 

While it cannot be deduced why ZS is unable to answer correctly without CoT, the PS offers some insights to identify what the model is lacking. For GPT-3.5-Turbo, out of the 29 questions that the PS failed to answer, 18 were due to attribute mapping errors, 8 were due to logical or coding errors, and the remaining were due to errors in the understanding of data structure. For GPT-4, out of 19 incorrect responses, 10 were due to logical or coding errors, 6 were due to data structures, and 3 were due to attribute mapping errors. Figure \ref{fig:retrieval_mistakes} in the Appendix shows example prompts and responses from GPT-4 where it fails to retrieve information due to logical and attribute mapping errors.

Besides accuracy, we measure the consistency of results between ZS and PS techniques for both GPT-3.5-Turbo and GPT-4. Figure \ref{fig:consistency_code} shows the consistency of outputs from both techniques. The average consistency for GPT-3.5-Turbo is about 30\%, and 57\% for GPT-4, indicating that the results from these two techniques are widely different, more so in the case of complex questions.

\section{Conclusion}\label{conclusion}
This paper tests the ability of GPT-3.5-Turbo and GPT-4 to answer questions about the GTFS specification and to extract information from the GTFS feed using two benchmarks. The LLMs are tested \emph{without any} additional input,  including the GTFS documentation. Overall, GPT-4 outperforms GPT-3.5-Turbo across all categories in both benchmarks. Through multiple-choice questions (MCQs) and zero-shot (ZS) prompting, we tested understanding and identified the shortcomings of GTFS in different categories. Although GPT-3.5-Turbo and GPT-4 work fairly well among most categories, they fail to accurately map categorical attributes to their corresponding values. To overcome this, one has to explicitly give the mapping information as an additional context or fine-tune the LLM. Also, we document that ChatGPT is not robust to option substitution. For the GTFS retrieval benchmark, the ZS prompting fails in most cases. The program synthesis (PS) works reasonably well for simple questions but struggles with hallucinations for complex queries. The PS output is significantly different from that of ZS as it may elicit a chain of thought. Besides, unlike ZS, PS can be used for asking generalized questions that are independent of a specific GTFS feed, thereby eliminating the bottleneck of context length. 



This paper demonstrates the capabilities of LLMs in comprehending GTFS data and utilizing this understanding to extract information from natural language prompts. The implications of this study may one day benefit academics, planners, local leaders, journalists, and anyone interested in parsing fine-grained transit data---particularly those with limited coding expertise. While LLMs have only recently become widely available, the study shows they can correctly answer a substantial number of detailed questions about GTFS. This raises the possibility that, in the future, LLM's may somewhat democratize transit data extraction and analysis. This could enable more informed community input and allow even agencies without staff coding expertise to perform queries that currently require programming.

The paper also demonstrates a systematic approach for evaluating and benchmarking LLM's comprehension of transit data. The set of questionnaires generated for this paper could also help with benchmarking other LLMs on GTFS understanding and information retrieval.

Currently, this paper works on a closed-source LLM. Therefore, the only improvement can be brought through in-context learning(i.e. prompt-based learning). Within in-context learning, breaking down the question into simpler steps could lead to better results.  The hallucinations with program synthesis are a result of misalignment in ``Attribute Mapping'', ``Data Structures'', and logical errors in coding. The GPT-3.5-Turbo and GPT-4 API versions do not have a code interpreter and are prone to coding errors. Using a built-in code interpreter could help overcome coding errors and also elicit thinking in the LLM. The usage of open-source LLMs and training the LLMs using supervised fine-tuning (SFT) and reinforcement learning with human feedback(RLHF) still need to be explored. 


\section{Data Availability}\label{data_avail}
Both `GTFS Understanding' and `GTFS Retrieval' benchmarks along with the filtered GTFS data and questionnaire used in this paper are available at \url{https://github.com/UTEL-UIUC/GTFS_LLM}.

\section{Acknowledgements}

This paper was initially drafted by the authors, and subsequently, certain sections were refined by \emph{gpt-3.5-turbo} and \emph{text-davinci-002} to enhance clarity and improve the overall eloquence of the writing.

\section{Author Contributions}
The authors confirm their contribution to the paper as follows: study conception and design: S. Devunuri; data collection:  S. Devunuri, S.Qiam; analysis and interpretation of results: S. Devunuri, S.Qiam, L. Lehe; draft manuscript preparation: S. Devunuri, S.Qiam, L. Lehe. All authors reviewed the results and approved the final version of the manuscript.


\newpage
\bibliography{sn-bibliography}

\newpage
\section*{Appendix}
\begin{figure}[!h]
    \centering
    \begin{subfigure}{\textwidth}
        \import{}{figures/categorical.tex}
        \caption{Zero-Shot (ZS) Prompt}
        \label{fig:categorical_map_ZS}
    \end{subfigure}
    \begin{subfigure}{\textwidth}
        \import{}{figures/categorical_cot.tex}
        \caption{Zero-shot with Chain of Thought (CoT) prompt}
        \label{fig:categorical_map_CoT}
    \end{subfigure}
    \caption{GPT-3.5-Turbo makes mistakes with `Categorical Mapping'. The difference between ZS \& CoT is the single instruction change highlighted in the system prompt. The `route\_type' for the bus is `3' according to the GTFS documentaion}
    \label{fig:categorical_map}
\end{figure}

\begin{figure}[!h]
    \centering
    \begin{subfigure}{\textwidth}
        \import{}{figures/retrieval_mistakes_1.tex}
        \caption{Example GTFS Retrieval question where the GPT-4 makes a logical mistake. The GPT-4 initially performs the merge of data frames correctly. However, the grouping is performed on `route\_type' alone instead of `route\_type' and `trip\_id'}
        \label{fig:retrieval_mistake1}
    \end{subfigure}
    \begin{subfigure}{\textwidth}
        \import{}{figures/retrieval_mistakes_2.tex}
        \caption{Example GTFS Retrieval question where the GPT-4 makes an `Attribute Mapping' error. The GPT-4 assumes that the attribute `shape\_dist\_traveled' is present in `trips.txt'. The `shape\_dist\_traveled' attribute is part of `stop\_times.txt'}
        \label{fig:retrieval_mistake2}
    \end{subfigure}
    \caption{Some examples where GPT-4 makes mistakes with GTFS Information Retrieval. The ``System" prompt is the same as the one in Figure \ref{fig:few-shot}}
    \label{fig:retrieval_mistakes}
\end{figure}

\end{document}